# UCMEXPORTER: Supporting scenario transformations from Use Case Maps

*Daniel Amyot, Ali Echihabi, and Yong He*
SITE, University of Ottawa
800 King Edward
Ottawa, Ontario, Canada, K1N 6N5
{damyot | u2090356 | yonghe}@site.uottawa.ca

**ABSTRACT**
The Use Case Maps (UCM) scenario notation is applicable to many requirements engineering activities. However, other scenario notations, such as Message Sequence Charts (MSC) and UML Sequence Diagrams (SD), have shown to be better suited for detailed design. In order to use the notation that is best appropriate for each phase in an efficient manner, a mechanism has to be devised to automatically transfer the knowledge acquired during the requirements analysis phase (using UCM) to the design phase (using MSC or SD). This paper introduces UCMEXPORTER, a new tool that implements such a mechanism and reduces the gap between high-level requirements and detailed design. UCMEXPORTER automatically transforms individual UCM scenarios to UML Sequence Diagrams, MSC scenarios, and even TTCN-3 test skeletons. We highlight the current capabilities of the tool as well as architectural solutions addressing the main challenges faced during such transformation, including the handling of concurrent scenario paths, the generation of customized messages, and tool interoperability.

## 1. INTRODUCTION

Many software development processes make use of multiple languages and notations. Often different phases involve different notations or languages, and those that are suited in one phase may not be so appropriate in another phase. For instance, some notations focus on descriptions of operations, services, or workflows independently of supporting platforms, whereas others focus on detailed refinements where distributed components, communication mechanisms, and platform-specific information are taken into consideration. Ensuring smooth transitions from one phase to the next as well as ensuring consistency between different views quickly become challenging.

Automated transformations are usually very beneficial in such context because they enable one to reuse information in one phase or view into another one, they help maintain some degree of consistency, and they can accelerate the development. Such transformations are at the basis of emerging approaches such as OMG's Model Driven Architecture (MDA) [19]. However, they also have their share of issues, including the formal representation of some of the concepts, the conceptual mapping from one notation to the next, the addition of information during refinement activities, and tool interoperability.

There are many success stories where scenario notations were used to design, validate, and reason about complex software and distributed systems (including telecommunication systems [3]). Among other notations, Use Case Maps (UCMs) [4] and UML Activity Diagrams (ADs) [18] conveniently express services and operational requirements, whereas Message Sequence Charts (MSCs) [9] and UML Sequence Diagrams (SDs) are well suited for detailed design. Being able to automate refinement transformations from UCMs/ADs to MSCs/SDs would be beneficial to a large community of developers.

This paper presents a tool, named UCMEXPORTER, for the automatic transformation of Use Case Map scenarios to Message Sequence Charts, UML 1.x Sequence Diagrams, and TTCN-3 test skeletons [10]. As shown in Fig. 1, UCMEXPORTER takes as input individual scenarios (in XML [26]) generated from the UCMNAV tool [25], and outputs:

- MSCs in Z.120 textual form, readable by Telelogic TAU SDL Suite [21];
- Sequence Diagrams in XML Metadata Interchange (XMI) [17], readable by Rational Rose [8];
- and textual TTCN-3 tests, readable by TTthree [22].

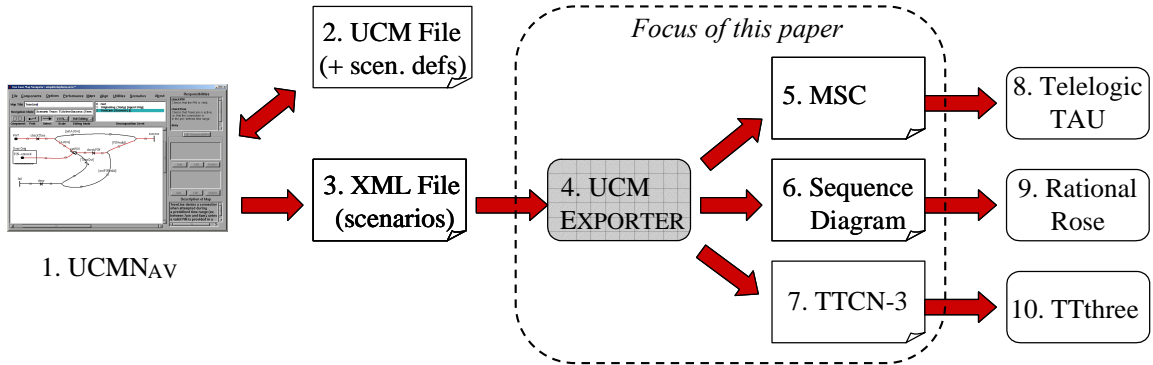

**Fig. 1** Usage of UCMEXPORTER.

Tools that generate individual scenarios (and other representations) from UCMs currently exist and are briefly discussed in section 2. Some of the most interesting challenges in transforming UCM-based scenarios to more detailed ones are covered in section 0, while section 4 presents our new solutions and the UCMEXPORTER architecture. A discussion of related work and conclusions then follow.

## 2. BACKGROUND AND PREVIOUS WORK

### 2.1 Use Case Maps

Message Sequence Charts and UML Sequence Diagrams and are famous in the distributed system community. However, Use Case Maps are less known and are hence briefly introduced here. The UCM notation is being standardized as part of the User Requirements Notation in the Z.150 series of ITU-T Recommendations [1][11][12]. UCM graphical models describe service requirements and high-level designs with causally linked responsibilities, superimposed on structures of components. UCM *responsibilities* are scenario activities representing something to be performed (operation, action, task, function, etc.). Responsibilities can potentially be allocated to *components*, which are generic and abstract enough to represent software entities (e.g. objects, processes, databases, or servers) as well as non-software entities (e.g. actors or hardware resources).

The four UCMs in Fig. 2 illustrate some of the notation's basic constructs. This brief example describes call request scenarios involving originating and terminating users and their agents. Start points capture triggering events and/or preconditions whereas end points represent resulting events and/or postconditions. UCM paths show the causal relationships among these points as well as among responsibilities. OR-forks and OR-joins represent alternative paths that split and merge, whereas AND-fork and AND-join capture concurrent paths that fork and synchronize. Alternatives can be guarded by conditions. Timers show points on a path that must be triggered in a timely fashion or else lead to the selection of a timeout path. Responsibilities can be allocated to generic components, represented as rectangles. The diamonds, called stubs, are containers for sub-maps called plug-ins. The root map in Fig. 2a contains two static stubs, which contain the maps in Fig. 2b and Fig. 2c. The stub in Fig. 2b is dynamic (shown with dashed lines) because it contains multiple plug-ins, one of which is shown in Fig. 2d. This stub could contain other plug-ins, which would describe different features that a user could subscribe to (not shown here). The selection among these plug-ins is performed according to a selection policy, which describes preconditions associated to each plug-in. A binding relationship specifies how start/end points in plug-ins are connected to stubs' IN/OUT segments.

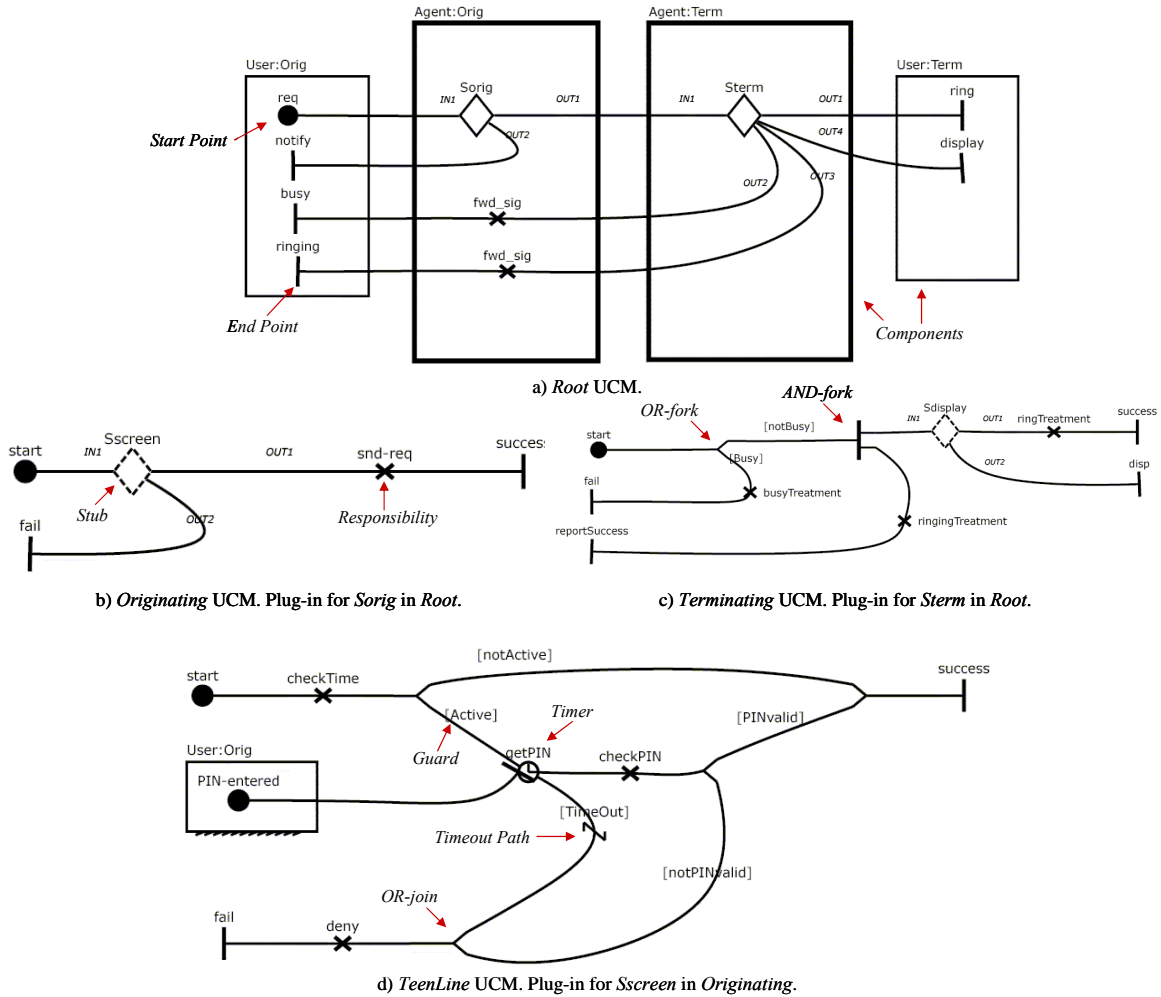

**Fig. 2** Sample UCM: Call Request in an agent-based telephone system with features

UCMs are useful for describing multiple scenarios abstractly in a single, integrated view. This promotes the understanding and reasoning about the system as a whole, as well as the early detection of conflicting or inconsistent scenarios [3]. However, for complex UCMs (e.g. with many alternative paths or with multiple levels of sub-maps), individual scenarios become difficult to recover. Yet, individual scenarios contribute greatly to the understanding of particular functionalities, and they also guide the definition of more detailed scenarios for the design phase, for validation test cases, and for documentation. Automated support for extracting individual scenarios from UCM requirements and for transforming them to design artifacts (in MSC or SD) or to test purposes (in TTCN-3 [10]) can help bridge the gap between requirements and design and validation. UCMEXPORTER is meant to be an extensible tool that addresses this problem using an appropriate mixture of technologies and where common problems (e.g. generation of messages and handling of concurrency) are solved in a central place, independently of the target language or notation. A particular attention was also given to interoperability with target CASE tools.

## 2.2 Existing UCM-based Transformations

The extraction of scenarios from UCMs is not without challenges. UCMs are very abstract in nature (e.g. they do not specify what messages are required to ensure the causal flow from one component to the next), and they can be constructed and combined in very flexible ways (e.g. parallel paths that are not well-nested can be constructed in a UCM graph). These problems were first tackled by Miga *et al.* in [15], and their solution was prototyped in the UCM Navigator (UCMNAV [25], now open-source), a multi-platform tool written in C++. This tool can highlight the UCM paths traversed according to *scenario definitions* (see Fig. 3).

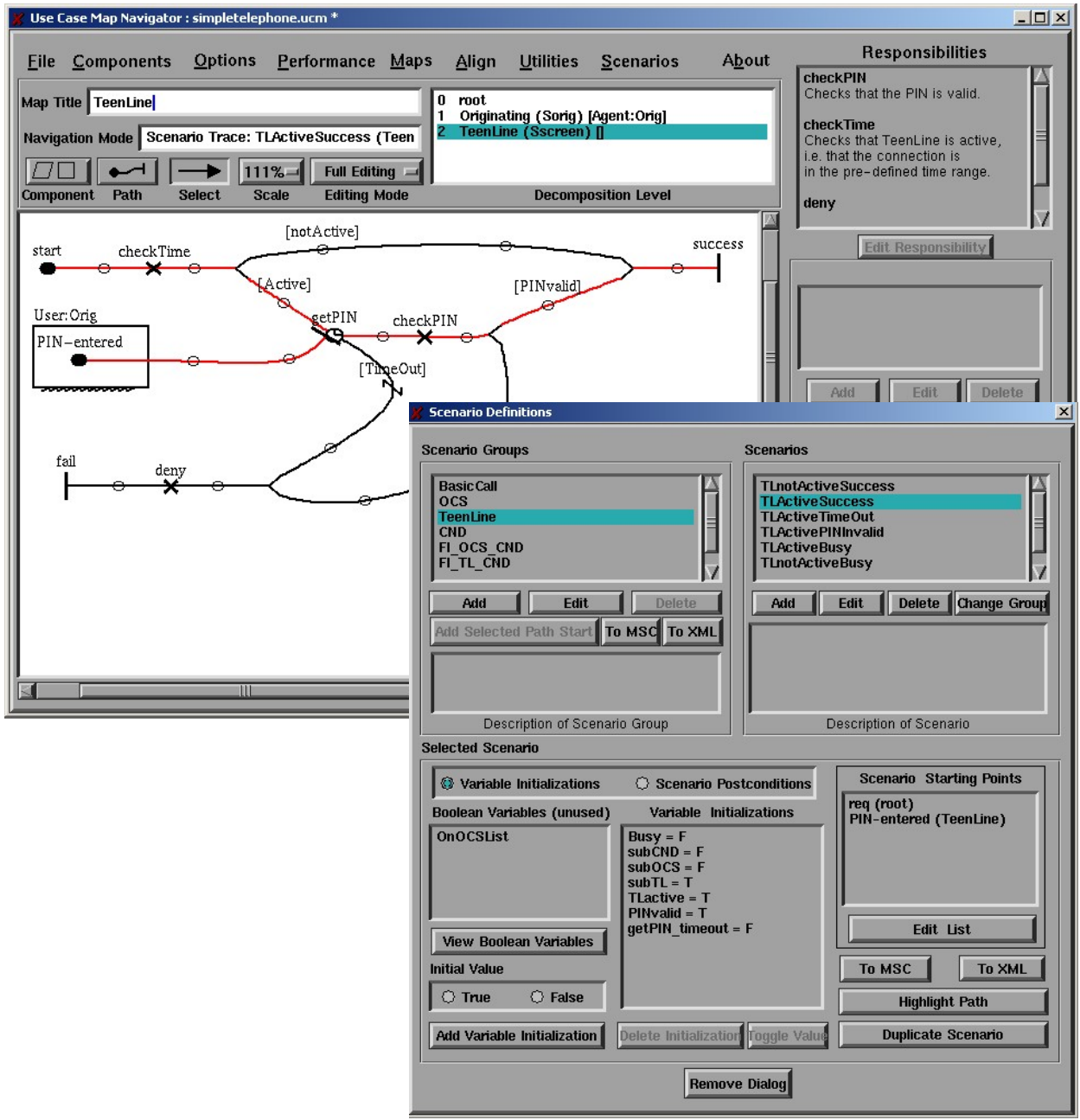

**Fig. 3** UCMNav and GUI for scenario definitions.

Each scenario definition provides the initial context for a named individual scenario in the form of:
- initial values for Boolean variables used in conditions for selection points such as OR-forks for alternative paths, timer conditions, and selection policies for dynamic stubs where various sub-maps can be selected;
- a set of UCM start points that are triggered initially;
- optional post-conditions to be satisfied once the scenario ends.

Scenario definitions are stored together with UCM descriptions in an XML file used by UCMNAV to load and save UCMs (see #2 in Fig. 1).

Miga *et al.* reused the path traversal mechanism based on scenario definitions to generate individual MSC scenarios directly. However, this approach is rather inflexible because the traversal of UCMs is intertwined with the generation of the target scenario representation (which makes the programming error-prone and difficult to maintain), and there is no way to adapt the traversal algorithm or modify the transformation to a different target scenario language. Nevertheless, this first prototype helped us gain a better understanding of the difficulties of traversing a UCM graph and of generating detailed scenarios.

In a recent paper [2], we introduced a new two-step approach to the generation of scenarios from UCMs that is more robust, flexible, and extensible. This approach decouples the traversal of UCMs from the generation of the target scenarios, and hence reduces the complexity of the algorithms and programs involved. UCMNAV was modified to support a new traversal algorithm initially proposed in the draft UCM standard [12].

In a nutshell, this algorithm uses a depth-first traversal of the graph that captures the UCMs' structure, which is backtracked to get the next available parallel branch of an AND-fork or the next start point (from the scenario definition) when it gets blocked at a stop point (timer, end point, or AND-Join). This approach treats these three types of stop points in a very similar way, which is simpler and more robust than the older algorithm used in [15]. Alternatives are resolved using the initial values of a user-defined set of Boolean variables, which can be modified in scenario responsibilities. Loops and hierarchical UCMs (with stubs and plug-in maps) are flattened during the traversal. UCM graphs may contain parallel constructs that are not well nested but the resulting individual scenarios are well nested; this means that, in specific situations, some of the concurrency found in the original UCM may not be preserved in the resulting scenario (however, no additional concurrency is added). These extracted scenarios need to be well nested so they can be converted to a linear representation such as MSC, SD, or test cases (see [2] for a more complete discussion).

The result of such traversal can be stored in an XML file (#3 in Fig. 1), whose structure is presented in the *Document Type Definition* (DTD) found in Fig. 4. Resulting scenarios are partial orders capturing the UCM nodes that have been traversed (in sequence and in parallel with `seq` and `par` elements). They also contain traceability links to the UCM components to which they belong and also to the source elements found in the original UCM file (#2 in Fig. 1).

The XML files compliant to this DTD become the basis for a second step, which consists of transforming the XML scenarios into specified target scenario language using *XSL Transformations* (XSLT, [27]). XSLT is usually well suited for XML-based transformations to various text-based representations, and many transformations can be used on the same XML scenario file to cover various target languages. An XSL (Extensible Stylesheet Language) stylesheet was developed to support the conversion towards MSCs and was used, with limited success, in [2]. MSCs were indeed generated automatically, and they were readable by Telelogic TAU, but:
- there were still situations where the MSCs generated handled concurrency incorrectly;

- XSLT is a somewhat limited language and the transformation (even if simpler) was still complex and unmanageable;
- some of the work was difficult to reuse in other contexts, e.g. for the generation of UML sequence diagrams in XMI.

```
<!ELEMENT scenarios (group)*>
<!ATTLIST scenarios
   date              CDATA   #REQUIRED
   ucm-file          CDATA   #REQUIRED
   design-name       CDATA   #IMPLIED
   ucm-design-version CDATA  #REQUIRED>

<!ELEMENT group (scenario)*>
<!ATTLIST group
   group-id          NMTOKEN #IMPLIED
   name              CDATA   #REQUIRED
   description       CDATA   #IMPLIED>

<!ELEMENT scenario (seq | par)>
<!ATTLIST scenario
   scenario-definition-id
                     NMTOKEN #IMPLIED
   name              CDATA   #REQUIRED
   description       CDATA   #IMPLIED>

<!ELEMENT seq (do | condition | par)*>

<!ELEMENT par (do | condition | seq)*>

<!ELEMENT do EMPTY>
<!-- WP_Enter: Reached a waiting place
 WP_Leave: Left the waiting place
 Connect_Start: Start point of a plug-in
 Connect_End: End point of a plug-in
 Trigger_End: Connected End point.  -->
<!ATTLIST do
   hyperedge-id      NMTOKEN #REQUIRED
   name              CDATA   #IMPLIED
   type   (Resp | Start | End_Point |
           WP_Enter | WP_Leave |
           Connect_Start | Connect_End |
           Trigger_End | Timer_Set |
           Timer_Reset | Timeout)
                             #REQUIRED
   description       CDATA   #IMPLIED
   component-name    CDATA   #IMPLIED
   component-role    CDATA   #IMPLIED
   component-id      NMTOKEN #IMPLIED >

<!ELEMENT condition EMPTY>
<!--"expression" is the Boolean
 expression used in the selected branch.
 "label" is the name describing the
 condition. -->
<!ATTLIST condition
   hyperedge-id      NMTOKEN #REQUIRED
   label             CDATA   #REQUIRED
   expression        CDATA   #IMPLIED >
```

**Fig. 4** DTD for scenarios extracted from UCMs.

In order to address these issues, we initiated the development of UCMEXPORTER, to be detailed in the next two sections.

## 3. CHALLENGES
The transformation of UCM-based scenarios leads to several interesting issues with the target languages and their tools. Four of the main challenges are briefly discussed in this section.

### 3.1 Message Synthesis
Use Case Maps focus on causal relationships between responsibilities and they do not include any explicit information about messages exchanged between components. However, in order to produce meaningful MSCs and SDs, such messages need to be synthesized so that causal flow is maintained across entities. For instance, if a UCM path has responsibility *doX* in component *C1* followed by *doY* in component *C2*, then, after *doX*, *C1* has to tell *C2* that *doY* should be performed. A synthetic message *m1* from *C1* to *C2* (noted *C1*—*m1*→*C2* in this paper) can achieve this. Such messages can indeed be created, but their management becomes complex when concurrent paths are involved. Moreover, message names have to be consistent across multiple detailed scenarios generated from the same causal UCM path.

### 3.2 Allowing for Customization
Customization can be done at many levels. A first one relates to how a UCM construct is mapped to a construct of the target language. For instance, UCM start points and end points can be mapped to mes-sages or to actions in a MSC, to self-messages or to messages from/to an environment lifeline in a UML Sequence Diagram, etc. Self-messages can also be used to capture responsibilities as if an object was calling one of its methods to perform some local activity.

A second level is the simple renaming of UCM names in the target representation (e.g. abstract UCM component names to lifeline names in a deployment UML sequence diagram).

A third level is concerned with how synthetic messages are refined into something more usable and meaningful for the design stage. For instance, the message *C1—m1→C2* from the previous section could be refined in many ways:
- simply but consistently renamed across individual scenarios: *C1—RequestDoY→C2*;
- parameters could be added: *C1—Request(doY)→C2*;
- be refined using a local protocol: *C1—Init→C2*, followed by *C2—Ack→C1,* followed by *C1—Request(doY)→C2*;
- be refined using newly introduced or intermediate components: *C1—Request→ORB*, followed by *ORB—Request→C2*.

Customization hence focuses on how very abstract scenarios and causal relationships can be refined into meaningful domain-specific or platform-specific artifacts. This is very useful when information on the target architecture is known (e.g. two components communicate via CORBA, SOAP, or SS7), or when multiple product architectures are targeted (e.g. in a product family).

### 3.3 Expressing Parallelism

UCMs are very good at capturing and combining multiple scenarios, even in the presence of concurrency. This information should be preserved when moving towards the design in order to maintain their integrity and to offer as much flexibility as possible to the designers. This is why the XML scenarios generated by UCMNAV are partial orders and support parallel scenario steps. However, not all target notations handle concurrency the same way. MSCs have a good support with co-regions and the `par` inline statement, whereas UML 1.x Sequence Diagrams do not really have first-class support (except through labeling). UML 2.0 SDs have more or less the same expressive power as MSCs, but the standard is not yet finalized and very few tools are available at this time.

Also, synthesizing messages for parallel UCM paths is no trivial task; the solutions proposed in [2][15] have limitations when concurrent paths involve many different components. UCM responsibilities can be allocated to any component (or to none, in which case a responsibility is said to be performed by the environment), and paths can be forked and synchronized in any way and in different components. Fig. 5 illustrates four interesting situations with partial UCMs and corresponding partial MSCs:
- In situation (a), parallel paths containing responsibilities located in different source components (A and C) synchronize and then cause another responsibility to be performed in a target component (B). Synthetic messages are required from all the source components (different from the target component) to the target component, before activities can start in the latter. Note that in this example, self-messages have been used to represent UCM responsibilities. Other user-defined mappings are possible (e.g. MSC actions are used to capture responsibilities in the next three situations), but this is a separate issue raised in section 3.2.
- In situation (b), there are multiple target components that need to be informed that they can proceed (B and C). This information needs to come from all the incoming parallel segments, whether they are from the same component (which is the case here with A) or from multiple components.

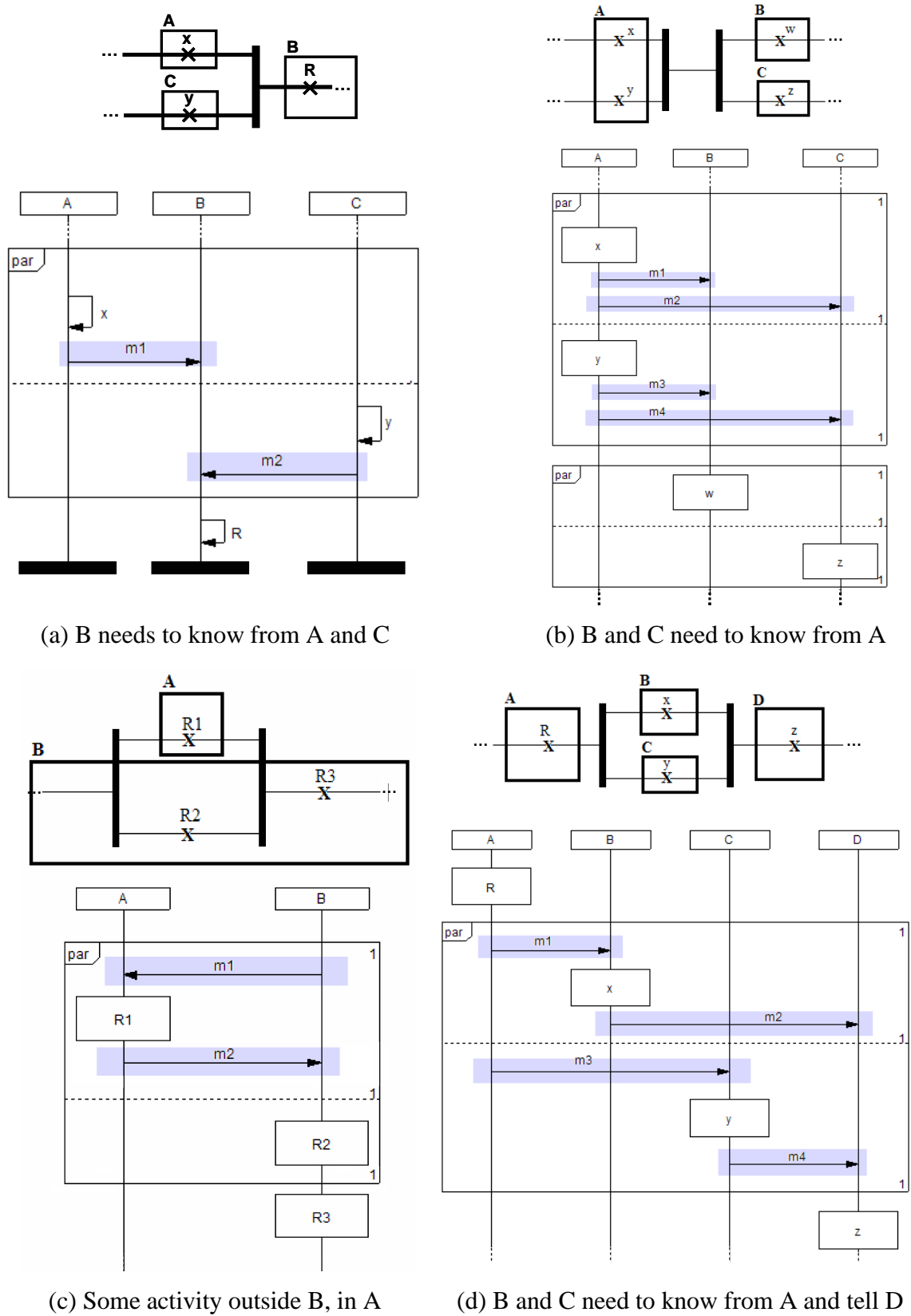

(a) B needs to know from A and C  
(b) B and C need to know from A  
(c) Some activity outside B, in A  
(d) B and C need to know from A and tell D

**Fig. 5** Four situations of partial mapping from a UCM to an MSC, with parallelism.

- In situation (c), paths can be forked in parallel and then synchronized to continue in the same component (B), yet some of the intermediate responsibilities may be located in different components (e.g. A) and appropriate communication needs to be established. Having *m1* and *m2* in the first parallel sub-sequence ensures maximum concurrency with the second subsequence (*R2*).
- Finally, in situation (d) the intermediate components (B and C) first need to receive confirmation that they can proceed (from A in this case), perform their responsibilities, and then tell the following components that they can proceed (D in this case). This is a generic case that handles sequences of the previous situations, which happen for long UCM paths involving many components.

Together, these four cases cover typical situations encountered in XML scenarios generated from UCM specifications. The relationship between UCM causality and messages is certainly not one-to-one: many synthetic messages are often required to "implement" what appears to be a simple causal relationship at the UCM level, and all these messages can further be customized as seen in the previous section.

### 3.4 Diagram Layout

Automatic layout of target design scenarios is both a language and a tool issue. For MSCs, the Z.120 textual notation does not contain layout information, and hence MSC tools are forced to provide an auto-layout function to import such models.

For SDs, the XMI standard (version 1.x, [17]) does not support layout information, but some popular extensions to XMI (e.g. from Unisys) can handle the positioning of elements in UML diagrams. XMI and such extensions are supported to various degrees in UML tools, often in incompatible ways. However, whereas many tools provide auto-layout for class diagrams, none really does the same for SDs and other types of UML diagrams, especially when they are expressed in XMI. We initially thought that generating SDs from UCMs would require simple modifications to the mechanism used to generate MSCs, however we could no longer rely on auto-layout mechanisms (not even by generating SDs in proprietary formats like Rational Rose's Petal/MDL files, which also require layout information) to produce *visual* SD models. The end of the next section will discuss our solution to this interoperability issue.

### 4. UCMEXPORTER

This section presents the main contribution of this paper, the UCMEXPORTER tool (and its architecture), as an answer to the challenges described in the previous section.

### 4.1 Architecture Overview

Fig. 6 gives a white-box view of the UCMEXPORTER box and of its context, as found in Fig. 1. The core of the logic is centralized in XMLInputTransform, a Java/XSLT set of transformations explained in detail in section 4.2. This produces a new XML file in UCMExporterXML format (containing messages), which can then easily and consistently be converted to various representations such as MSCs, SDs, and TTCN-3. Such conversions are mainly of syntactic nature (and were therefore implemented in XSLT), except for the generation of XMI code with layout information, which requires an additional step (discussed in section 4.3).

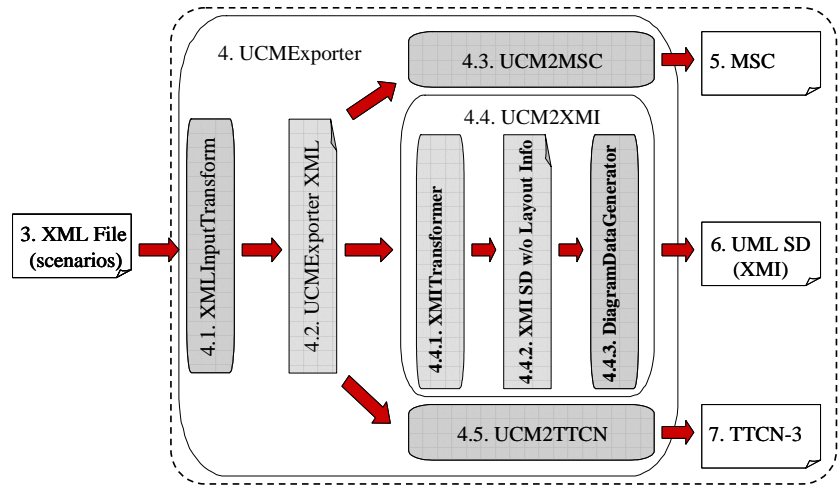

**Fig. 6** White box view of UCMEXPORTER.

## 4.2 XMLInputTransform

The goal of this function is to produce the additional data (i.e. list of component instances involved and intermediate messages synthesized) necessary for the generation of MSCs, SDs, and TTCN to become straightforward and implementable using simple XSL stylesheets. It is composed of a sequence of six transformations, shown in Fig. 7, and produces as output a new XML file compliant to the *UCMExporterXML* Document Type Definition. This DTD is backward compatible with the *scenario* DTD described in Fig. 4, but it extends the latter with an optional list of component instances (useful for declaring MSC instances and SD object lifelines) and optional messages found in sequences and parallel elements. These messages have the following XML format:

```
<!ELEMENT message EMPTY>
<!ATTLIST message
    id              NMTOKEN        #REQUIRED <!- Unique message ID -->
    name            CDATA          #REQUIRED
    source-id       CDATA          #REQUIRED <!- Source component instance -->
    destination-id  CDATA          #REQUIRED <!-- Target component instance -->
    is-task         (true | false) #REQUIRED <!- Comes from resp. -->
    is-timer        (true | false) #REQUIRED <!- Comes from timer -->
    timer-property  CDATA          #REQUIRED <!- Timer attribute -->
    last-ref        CDATA          #REQUIRED <!- Used by parallel rules -->
    description     CDATA          #IMPLIED
    para-label      CDATA          #REQUIRED <!- Label for parallel segments -->
    is-connector    (true | false) #REQUIRED <!- Message to be refined -->
    connnector-type CDATA          #IMPLIED  <!-- Used by parallel rules -->
>
```

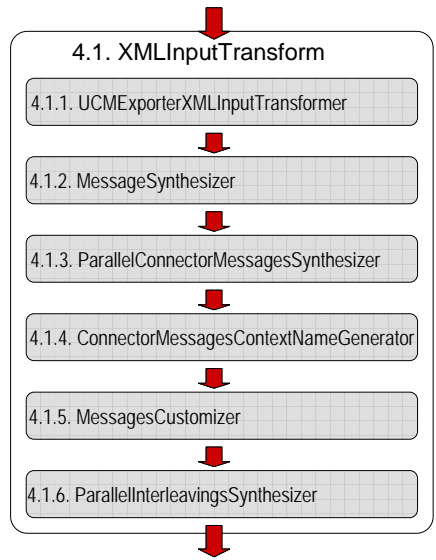

**Fig. 7** Steps followed during the transformation to XML scenarios with messages.

The six steps, which combine Java programs and XSL transformations, are:
- **UCMExporterXMLInputTransformer**: Parses the UCMNav XML input in order to identify all component definitions in the model as well as the instances participating in every scenario. Generates a UCMExporterXML file that includes this information at the beginning.
- **MessageSynthesizer**: Parses the output of the previous step and identifies the messages that are being sent in each interaction. It uses the instances and components defined in the previous step to identify the sender and receiver of each message. Several rules are applied to synthesize messages that preserve the causality. The messages are given dummy names. This step does not cover parallel-related messages because they require more complex rules for preserving the causality.
- **ParallelConnectorMessagesSynthesizer**: Applies four parallel connector rules to synthesize causality-preserving messages. These complementary rules handle the situations discussed in section 3.3:
    a) The last instance (or instances) before entering a parallel block;
    b) The instances that first execute inside the parallel block;
    c) The instances that last execute inside the parallel block;
    d) The first instance (or instances) after exiting the parallel block.

  This step uses XSL to determine the instances that last communicate before we enter a parallel block, the first that interact inside the parallel block, the last that interact in the parallel block, and the first that interact when we leave the parallel block. This information is again stored in an intermediate UCMExporterXML file (in the message/connector fields mentioned above). Then, a Java application takes this information and applies the rules for situations a, b, c, d. Existing rules could be tailored and new ones added locally without affecting the rest of the tool.
- **ConnectorMessagesContextNameGenerator**: All synthesized connectors (messages) in the previous steps are given dummy names. This step replaces those names with context-dependent and more descriptive names. For instance, messages *m2* and *m3* in Fig. 5(b) would be automatically renamed *did_x_do_z* and *did_y_do_w* respectively (see Fig. 8). This step also ensures that the same message name has the same meaning across a collection of scenarios generated from the same UCMs.
- **MessagesCustomizer**: Applies an optional XSL stylesheet provided by the user when he/she requires some customization of the transformation results. As explained in section 3.2, users could choose to rename various scenario elements, add parameters, refine messages with more complex and realistic protocols, introduce intermediate components in communications, etc. This is easily done with simple additional XSL rules, independently of the previous steps.
  – **ParallelInterleavingsSynthesizer**: Generates all possible interleavings in a parallel block and produces a set of sequential scenarios. This functionality is needed for target languages or tools which cannot accept scenarios that include parallelism. This is the case for Klocwork's MSC2SDL synthesizer, which takes sequential MSCs as input and produces a SDL formal specification out of them [7][14][21]. At this point, this functionality generates the simplest possibility (all parallel activities turned into one sequence), which is still sufficient to interoperate with Klocwork's tools, but it could be improved to support more complex interleaving schemes.

### 4.3 XSL Stylesheets and Code Generation

Three simple XSL stylesheets are currently available to transform a UCMExporterXML file using Apache's XSLTC engine [23]:
  – **UCM2MSC**: Generates a `.msc` file compliant to the Z.120 textual MSC notation, which can then be read and displayed by Telelogic TAU's MSC Editor [21] and by other tools such as Klocwork's MSC2SDL synthesizer [14].
  – **UCM2TTCN**: in a way similar to UCM2MSC, generates a `.ttcn` file compliant to the TTCN-3 syntax and executable by tools such as TTthree [22]. The XSL stylesheet is based on the work of Mulvihill [16], which was done prior to the creation of UCMEXPORTER. Hence, it does not take advantage of the presence of messages in its input XML file, and simply generates one monolithic test component. Evolving the XSL stylesheet to consider messages has been left for future work.
  – **UCM2XMI**: Again, a simple stylesheet is used to generate a UML Sequence Diagram in a `.xmi` file, using the standard XMI format (v. 1.x). Parallel constructs are transformed to sequential messages where the first parameter captures the concurrency information. For instance, everything labeled *P1.S1* is in parallel with everything tagged with the same *P* but with a different *S*, e.g. *P1.S2* and *P1.S3* (compare the MSC and SD in Fig. 8). Our labeling scheme also considers nested parallelism (e.g. *P1.S2.P2.S3.P1.S1*). However, as explained in section 3.4, no layout information is included by default, and the resulting diagram cannot be visualized as such on any UML tool.

If the user chooses to include diagram information, the *DiagramDataGenerator* function (see #4.4.3 in Fig. 6) will parse the standard XMI file and add minimal layout information so that it can be rendered properly. So far, we only have a *RationalRoseDataGenerator*, which is a Java program that adds Unisys-specific tags to the XMI file. Rational Rose uses a Unisys plug-in to import and render such XMI files. We selected this tool because it has undocumented features that provide some auto-layout when simple layout data is provided (e.g. horizontal positions of the lifelines). The order in which the messages are exchanged (or methods invoked) is important, but not their absolute vertical position. This has greatly simplified the computations required to provide such data. Other UML tools may or may not offer similar features, but this is hard to determine because many tools lack proper documentation regarding their support for XMI import/export.

Lastly, if the original XML file produced by UCMNAV contains multiple scenarios in the same file, then they will all be included in a single `.xmi` document as separate interactions (SD).

Fig. 8 presents detailed MSC and UML SD scenarios produced by UCMEXPORTER for one scenario supported by the sample UCM in Fig. 2. After a request (*req* in Fig. 2a), the plug-in in Fig. 2b is traversed. At stub *Sscreen*, the default plug-in is selected (this plug-in is not shown in Fig. 2, but it simply connects IN1 to OUT1 in *Sscreen*). Then, the *snd-req* is performed, and the flow exits at OUT1 after *SOrig*. The flow then continues in a different component, hence requiring a synthetic message to be produced (*did_snd_req_do_ringTreatment*). In this scenario, the terminating user is not busy, so two things happen in parallel: the terminating user will receive a *ring* signal, and the originating user will have a *ringing* signal. In this UML Sequence Diagram, the two parallel subsequences are named p1.s1 and p1.s2.

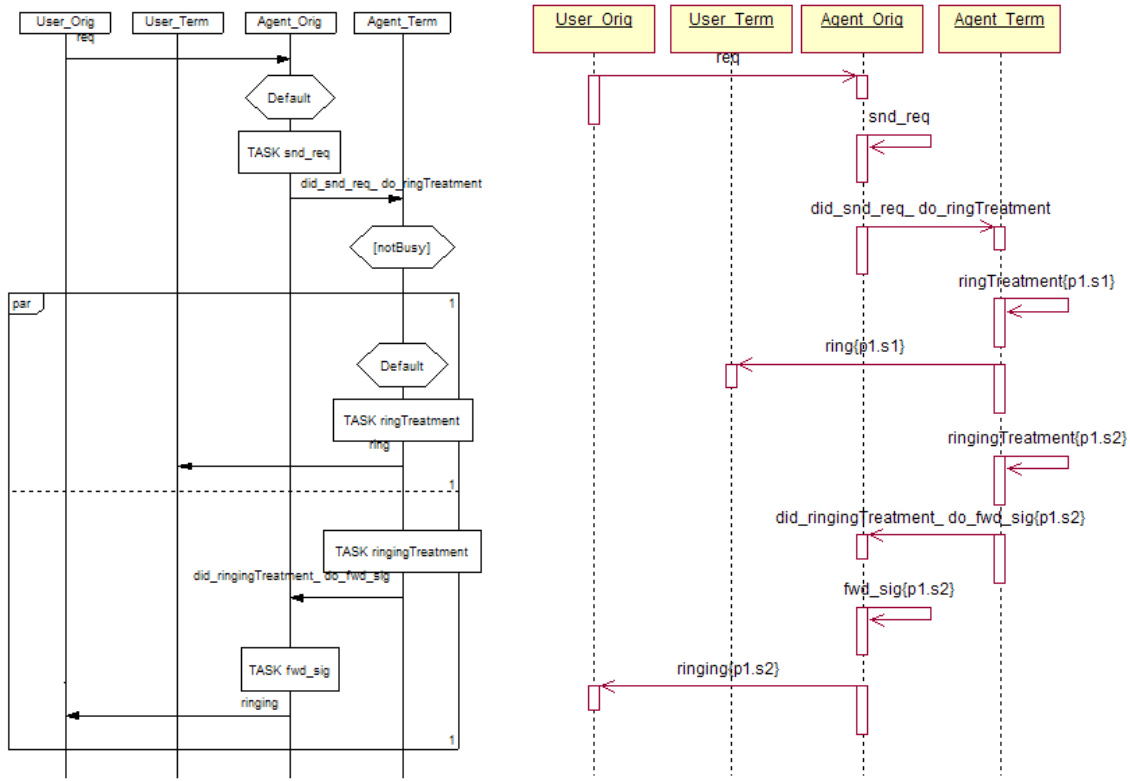

**Fig. 8** Sample MSC and SD produced by UCMEXPORTER from the same scenario.

Such transformation produces detailed but understandable scenarios. Without it, the reader would need to flip nine times through the different UCMs of Fig. 2 to follow the same scenario. Many more MSC/SD scenarios can be generated automatically in the same way from UCMs combined with scenario definitions.

### 4.4 UCMEXPORTER GUI
UCMEXPORTER offers a simple Java-based GUI (Fig. 9) to handle the selection of a desired transformation (currently to MSC, XMI, or TTCN-3) and various options such as the location of the source and target scenario files, the location of the basic and customizing generation rules in XSL, and the opportunity of selecting an appropriate way of adding layout information to XMI files (only one option, for Rational Rose, is available at the moment). A Web-based interface is also under construction.

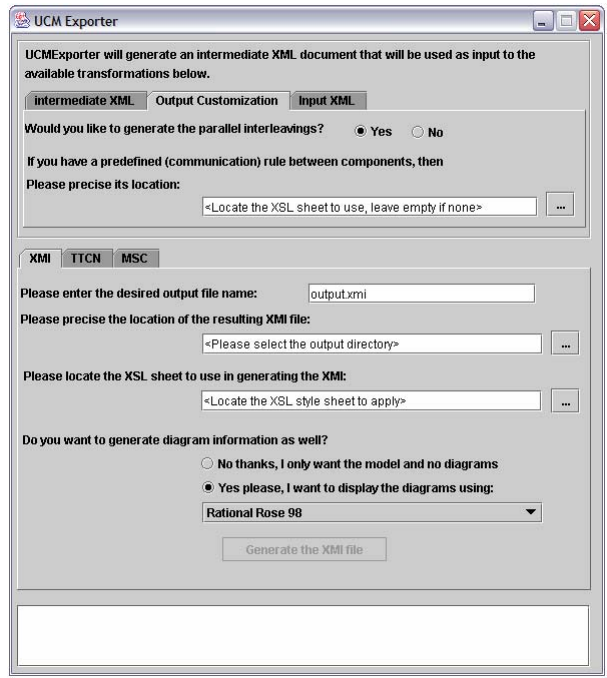

**Fig. 9** UCMEXPORTER GUI

## 5. DISCUSSION

UCM-based transformations are not a novel idea. In the past few years, several automated conversions have been proposed:
- Generation of Layered Queuing Networks, used in performance modeling. This was recently incorporated to UCMNAV [20]. This transformation requires the whole UCM graph to be considered, not just individual scenarios generated using scenario definitions and the UCM traversal algorithm.
- Guan explored the automated generation of LOTOS specifications from UCMs [6]. Her algorithm translates the whole UCM graph at once, without using scenario definitions or conditions that involve the global variables. This leads to a complete and executable prototype, but many incorrect scenarios (unfeasible paths caused by the non-evaluation of guarding conditions) can emerge in the resulting LOTOS model.
- He *et al.* recently used Miga's MSC-generation mechanism [15] to generate MSCs and then synthesize an SDL model, which is executable and amenable to formal validation and detection of problems such as unspecified receptions, deadlocks, livelocks, etc [7]. Many restrictions to the synthesizer used in that experiment (e.g. special naming conventions and no parallelism) can be overcome by providing simple custom XSL rules in UCMEXPORTER.

Our approach also shares some similarity with that of Gu and Petriu [5], who explored the XSLT-based transformations of UML activity/sequence diagrams to Layered Queuing Networks. Our generation process is less ambitious as it focuses on simpler scenarios, but it is more flexible because many target languages can be easily supported. We also use Java where it is the most beneficial, instead of having a complex solution based solely on XSLT.

Overall, the architecture of our tool offers many benefits:
- Transformations rules can be evolved, maintained, and corrected independently. For example, UCM2TTCN could be modified to take messages into consideration.

- Users can simply and efficiently customize some transformation aspects.
- New target languages (e.g. UML 2.0 SDs or Activity Diagrams) can easily be added via the addition of a new XSL file and slight modifications to the GUI.
- The intermediate steps combine all the work needed by the transformations (except the final code generation), avoiding redundancy and duplication of code. This will minimize the efforts required to maintain the tool.
- The separation of *XMITransformer* and *DiagramDataGenerator* enables the creation of new diagram data generators for target UML tools other than Rational Rose, without modifying the generation of the standard XMI.
- The use of Java and XSLT makes UCMEXPORTER a platform-independent tool.

## 6. CONCLUSIONS

This paper has described the new UCMEXPORTER tool, used to transform abstract UCM scenarios into detailed MSC and UML SD scenarios, and into TTCN-3 test skeletons. Section 0 has described some of the main challenges faced while doing such transformations, whereas section 4 presented our solutions and tool architecture, selected to satisfy goals such as usage flexibility, and tool interoperability, maintainability, and extensibility. This tool contributes to narrowing the gap between requirements definition/analysis phases and design/testing phases by automating the transfer of important knowledge (scenarios) and by reducing the time and effort to be invested in doing that transfer. This is in line with the spirit of transformations between platform-independent models and platform-specific models, as suggested by MDA [19].

With UCMEXPORTER, we now possess the infrastructure required to prototype and evaluate many options and rules that could potentially be used in scenario-driven requirements analysis, design, and testing. In particular, we plan to improve the TTCN-3 generation and eventually support UML 2.0 Sequence Diagrams (similar to MSC) when the UML 2.0 XMI representation and tools become available. UCMEXPORTER is also an open source project to which people can contribute. We hope that future usage of this tool will influence the way we use UCMs and other scenario notations, and we expect to be in a position to contribute to emerging methodological standards such as ITU-T's Z.153, which discusses relationships between the User Requirements Notation (including UCM) and related description languages [13].

**Acknowledgments**
This work was supported by the Natural Science an Engineering Research Council of Canada as part of the strategic project: Requirements-Driven Development of Distributed Systems.